# Interparticle Coupling Effects of Two Quantum Dots System on the Transport Properties of a Single Plasmon

Nam-Chol Kim,[1,2,*] Myong-Chol Ko,[1] Song-Il Choe,[1] Chol-Jong Jang,[1] Gwang-Jin Kim,[1] Zhong-Hua Hao,[2] Jian-Bo Li,[3] and Qu-Quan Wang[2,4,*]

[1]Department of Physics, **Kim Il Sung** University, Pyongyang, DPR of Korea
[2]School of Physics and Technology, Wuhan University, Wuhan 430072, China
[3]Institute of Mathematics and Physics, Central South University of Forestry and Technology, Changsha 410004, China
[4]The Institute for Advanced Studies, Wuhan University, Wuhan 430072, China
*ryongnam10@yahoo.com, qqwang@whu.edu.cn

**Abstract:** Transport properties of a single plasmon interacting with two quantum dots (QDs) system coupled to one-dimensional surface plasmonic waveguide are investigated theoretically via the real-space approach. We mainly focus on the coupling effects of the two QDs on the transmission properties of a single incident plasmon. We demonstrated that switching of a single plasmon can be achieved by controlling the interparticle distance, the interparticle coupling strength, and the QD-waveguide coupling strength, as well as spectral detuning. We also showed that the coupling between the continuum excitations and the discrete excitations results in the Fano-type transmission spectrum. The transport properties of a single plasmon interacting with such a two direct coupled QDs system could find the applications in the design of plasmonic nanodevices, such as single photon switching and nanomirrors, and in quantum information processing.

## 1. Introduction

[*] Electronic mail: ryongnam10@yahoo.com, qqwang@whu.edu.cn

A single-photon transistor is a device where the propagation of a single signal photon can be controlled by the presence or absence of a single gate photon [1]. Such a nonlinear device is essential to many emerging potential technologies, such as optical communication [2], quantum computer [3], and quantum information processing [4, 5]. Photons can be regarded as ideal carriers of quantum information, therefore manipulating photons could have important applications in quantum computation and quantum information technology [6-8]. However, photons rarely interact with each other, thus we have to explore the ways how to control the photons with the photon-atom interaction. On the other hand, the coupling between a photon and an atom in the vacuum is usually very weak. Anyway, we can modify the coupling strength by changing the environment of the vacuums by Purcell effect [9]. Strong coupling of the interaction between a single photon and atoms could be achieved by confining the single photon in reduced dimensions such as in one dimensional (1D) photonic waveguide with transverse cross sections on the order of a wavelength square [10]. Recently, as a scheme to achieve strong coupling between light and an emitter, surface plasmons (SPs), which are propagating electromagnetic modes confined to the surface of a conductor-dielectric interface, have attracted intensive interests [11-17].

Especially, the transport properties of a single photon (plasmon) interacting with quantum emitters has been investigated in the real-space approach [18-20]. It is useful to resort to real-space approach for determination of the response to the single injected photon, which is particularly convenient for discussing photon transport from one space-time point to another one and makes no assumptions on temporal behaviors of the constituents of the system. Therefore, a single photon transport based on the real-space method has a period of explosive growth, including the transport properties of the single photon (plasmon) interacting with quantum emitters which have various structures, for example, such as two-level emitter [21], three-level emitter [22,23], and multiparticle emitters[24-30]. However, in the previous researches concerned on the transport properties of a single plasmon interacting with multiparticle quantum emitters, the interactions between individual quantum emitters have not been considered yet. Motivated by these considerations, we investigate the transport properties of a single plasmon interacting with two emitters, coupled to 1D surface plasmonic waveguide, where the emitters could be quantum dots (QDs). In the present paper, we mainly focus on the influence of the coupling effects of the two QDs on the transport properties of a single incident plasmon interacting with such a coupled two QDs system.

## 2. The model of the system and the solutions

We consider the transmission properties of an incident single plasmon interacting with a two QDs system coupled to a 1D plasmonic waveguide which can be a metal nanowire (MNW), as shown in Fig. 1. We model the QDs as spherical semiconductors with a dipole located at the center of each. We treat each QD as effective two level quantum systems with transition energies $\hbar\Omega_1$ and $\hbar\Omega_2$, transition dipole moments $\mathbf{D}_1$ and $\mathbf{D}_2$, respectively. The QDs interact with each other with dipole-dipole interaction via Coulomb force. We will treat the problem in real space, which is particularly convenient for discussing photon transport properties from one space-time point to another one [18]. Also, the treatment is exact and makes no assumptions on temporal behaviors of the constituents of the system.

Under rotating wave approximation, the Hamiltonian of the whole system consisting of a single incident plasmon and two QDs coupled 1D plasmonic waveguide can be given by [31],

$$H/\hbar = \sum_{j=1}^{2}\left[(\Omega_j - i\Gamma'_j/2)\hat{\sigma}_j^+\hat{\sigma}_j\right] + \sum_k \omega_k a_k^+ a_k + \sum_{j=1}^{2}\sum_{\substack{i=1\\i\neq j}}^{2} f_{ji}\hat{\sigma}_j^+\hat{\sigma}_i + \sum_{j=1}^{2}\sum_k g_j\left(a_k^+\hat{\sigma}_i + a_k\hat{\sigma}_j^+\right) \quad (1)$$

We can transform the Hamiltonian of the system in real space as [18, 23]

$$\begin{aligned}H/\hbar =& \sum_{j=1}^{2}\left[(\Omega_j - i\Gamma'_j/2)\hat{\sigma}_j^+\hat{\sigma}_j\right] + iv_g\int_{-\infty}^{\infty}dz\left[a_l^+(z)\partial_z a_l(z) - a_r^+(z)\partial_z a_r(z)\right] \\ &+ \sum_{j=1}^{2}\sum_{i=1(i\neq j)}^{2} f_{ji}\hat{\sigma}_j^+\hat{\sigma}_i + \sum_{j=1}^{2}g_j\left\{\left[a_r^+(z_j) + a_l^+(z_j)\right]\hat{\sigma}_j + \left[a_r(z_j) + a_l(z_j)\right]\hat{\sigma}_j^+\right\},\end{aligned} \quad (2)$$

where $\Omega_j$ are the eigenfrequencies of the $j$th QD, respectively, $\omega_k$ is the frequency of the surface plasmon with wavevector $k$ ($\omega_k = v_g |k|$). $\hat{\sigma}_j = |g\rangle_{jj}\langle e|$ ($\hat{\sigma}_j^+ = |e\rangle_{jj}\langle g|$) is the lowing (raising) operators of the $j$th QD, $a^+_r(z_j)$ ($a^+_l(z_j)$) is the bosonic operator creating a right-going (left-going) plasmon at position $z_j$ of the $j$ th QD. $v_g$ is the group velocity of the surface plasmon, the non-Hermitian term in $H$ describes the decay of state $\omega_0^{(j)}$ at a rate $\Gamma'_j$ into all other possible channels. $g_j = (2\pi\hbar/\omega_k)^{1/2}\Omega_j \mathbf{D}_j\cdot\mathbf{e}_k$ is the coupling constant of the $j$ th QD with surface plasmon, $\mathbf{e}_k$ is the polarization unit vector of the surface plasmon and $f_{ji}$ is the interparticle coupling strength between the two QDs[18]. We assume that a single plasmon is incident from the left with energy $E_k = \hbar\omega_k$, then the eigenstate of the system, defined by $H|\psi_k\rangle = E_k|\psi_k\rangle$, can be found in the form

$$|\psi_k\rangle = \int dz \left[\phi^+_{k,r}(z)a^+_r(z) + \phi^+_{k,l}(z)a^+_l(z)\right]|0, g_1, g_1\rangle + e_k^{(1)}|0, e_1, g_2\rangle + e_k^{(2)}|0, g_1, e_2\rangle. \quad (3)$$

Here $|\bullet,\bullet,\bullet\rangle$ denotes the state $|SP, QD-1, QD-2\rangle$, for example, $|0, e_1, g_2\rangle$ denotes the vacuum state with zero plasmon, and $e_i$ or $g_i$ labels the excited or ground state of the QD-$i$, respectively, and $e_k^{(j)}$ is the probability amplitude of the $j$ th QD in the excited state. $\Phi^+_{k,r}(z)$ ($\Phi^+_{k,l}(z)$) is the wavefunction of a right-going (a left-going) plasmon at position $z$. For a single plasmon incident from the left, the mode functions $\Phi^+_{k,r}(z)$ and $\Phi^+_{k,l}(z)$ take the forms $\phi^+_{k,r}(z<0) = e^{ikz}$, $\phi^+_{k,r}(0<z<L) = t_1 e^{ik(z-l)}$, $\phi^+_{k,r}(z>L) = t_2 e^{ik(z-2l)}$, $\phi^+_{k,l}(z<0) = r_1 e^{-ikz}$, $\phi^+_{k,l}(0<z<l) = r_2 e^{-ik(z-l)}$, and $\phi^+_{k,l}(z>l) = 0$, respectively, where $j = 1, 2$ and $l$ is the spacing between the two QDs. Here $t_j$ and $r_j$ are the transmission and reflection amplitudes at the place $z_j$, respectively. By substituting Eq. (3) into $H|\psi_k\rangle = E_k|\psi_k\rangle$, we obtain a set of equations as: $r_2 e^{ikl} - r_1 - \frac{ig_1}{\upsilon_g} e_k^{(1)} = 0$,

$t_1 e^{-ikl} - t_0 + \frac{ig_1}{\upsilon_g} e_k^{(1)} = 0$, $r_3 e^{ikl} - r_2 - \frac{ig_2}{\upsilon_g} e_k^{(2)} = 0$, $t_0 + r_1 + \frac{(\Delta_k^{(1)} - i\Gamma'_1/2)}{g_1} e_k^{(1)} + \frac{f_{12}}{g_1} \cdot e_k^{(2)} = 0$,

$t_2 e^{-ikl} - t_1 + \frac{ig_2}{\upsilon_g} e_k^{(2)} = 0$, $t_1 + r_2 + \frac{(\Delta_k^{(2)} - i\Gamma'_2/2)}{g_2} e_k^{(2)} + \frac{f_{12}}{g_2} \cdot e_k^{(1)} = 0$. By taking the boundary conditions of the mode functions $t_{j-1} + r_j = t_j e^{-ikl} + r_{j+1} e^{ikl}$, where $t_0 = 1$, $r_3 = 0$ ($j = 1, 2$) into account in the set of the above equations, we obtain the transmission and the reflection amplitudes, respectively, as follows

$$t_2 = \frac{e^{ikl}\upsilon_g\left[(-1+e^{i2kl})fg^2 + ie^{ikl}(f^2 - \Delta_1\Delta_2)\upsilon_g\right]}{-i(-1+e^{i2kl})g^4 + g^2(2e^{ikl}f - \Delta_1 - \Delta_2)\upsilon_g + i(f^2 - \Delta_1\Delta_2)\upsilon_g^2}, \quad (4\text{-a})$$

$$r_1 = \frac{ig^2\left[(-1+e^{i2kl})g^2 - i(-2e^{ikl}f + e^{i2kl}\Delta_1 + \Delta_2)\upsilon_g\right]}{-i(-1+e^{i2kl})g^4 + g^2(2e^{ikl}f - \Delta_1 - \Delta_2)\upsilon_g + i(f^2 - \Delta_1\Delta_2)\upsilon_g^2}, \quad (4\text{-b})$$

$$e_k^{(1)} = -\frac{g\upsilon_g\left[\left(-1+e^{i2kl}\right)g^2 + i\left(e^{ikl}f - \Delta_2\right)\upsilon_g\right]}{-i\left(-1+e^{i2kl}\right)g^4 + g^2\left(2e^{ikl}f - \Delta_1 - \Delta_2\right)\upsilon_g + i\left(f^2 - \Delta_1\Delta_2\right)\upsilon_g^2}, \quad \text{(4-c)}$$

$$e_k^{(2)} = -\frac{g\left(f - e^{ikl}\Delta_1\right)\upsilon_g^2}{-i\left(-1+e^{i2kl}\right)g^4 + g^2\left(2e^{ikl}f - \Delta_1 - \Delta_2\right)\upsilon_g + i\left(f^2 - \Delta_1\Delta_2\right)\upsilon_g^2}, \quad \text{(4-d)}$$

where we set $g_1 = g_2 = g$, $f_{12} = f_{21} = f$, $\Delta_k^{(j)} = \Omega_j - \omega_k$, $\Delta_j \equiv \Delta_k^{(j)} - i\Gamma_j'/2$ ($j = 1, 2$)), which have the same unit as frequency[18].

### 3. Theoretical Analysis and Numerical Results

The transport properties of a single plasmon in the long time limit can be characterized by the transmission (reflection) coefficient, $T_2 = |t_2|^2$ ($R_1 = |r_1|^2$). In all our calculations, we suppose that $\Gamma_1' = \Gamma_2' = 0$ and $J_1 = J_2 = J$, where $J_j = g_j^2/\upsilon_g$. First of all, we consider the case where the transition energies of the two QDs are equal to each other, $\Omega_1 = \Omega_2$. Fig. 2 shows the transmission coefficients versus the detuning for various interparticle distances ($l$), interparticle coupling strength ($f$), and the QD-waveguide coupling strength ($J$). Figs. 2(a) and 2(b) show the transmission coefficients versus the detuning $\Delta$ for the fixed interparticle distance $l=\lambda/4$ with the QD-waveguide coupling strength $J=0.1$ and 1, respectively. Figs 2(c) and 2(d) show the transmission coefficients versus the detuning $\Delta$ for the fixed interparticle distance $l=\lambda/20$ with the QD-waveguide coupling strength $J=0.1$ and 1, respectively. We can find that the transmission spectrum of a single incident plasmon is symmetric with respect to the zero detuning when the interparticle distances $l=n\lambda/4$ ($n=0,1,2,\cdots$)[Figs. 2(a) and 2(b)], otherwise asymmetric[Figs. 2(c) and 2(d)]. What is more interesting is that in all the above cases there appears the double complete reflection peak, which is quite different from the case that the two QDs have no interactions each other shown in Ref [23]. Furthermore, the spacing between the complete reflection peaks becomes wider, as the interparticle coupling strength becomes stronger. We also found that the transmission is enhanced, as the QD-waveguide coupling strength becomes small, as shown in Figs. 2(a) and 2(c). One can also find that none of the widths of the transmission peaks of the case $l=n\lambda/4$ is as sharp as that of the case $l\neq n\lambda/4$, which implies that the interparticle distance $l\neq n\lambda/4$ is more useful for a single plasmon switching. It should be noted that the width of the complete reflection peak of a single plasmon becomes wider by properly setting the parameters, for example, $f=0.1$, $J=1$, $kl = 0.5\pi$, which is quite an interesting result. In those cases of a single emitter[18], there appears a complete reflection peak only in a specific frequency, i. e., resonant frequency, which gives rise to the physical difficulties in practical applications, because the optical pulse controlled is actually a superposition of the plane wave with different frequencies where the off-resonant components could deviate from the complete reflection dramatically. Our result shows that properly adjusting physical parameters such as the QD-waveguide coupling strength, the interparticle coupling strength and distances results in broadening of the band of complete reflection, which could find practical applications. We also found that the position of the complete reflection peak can be controlled by adjusting the interparticle coupling strength, as shown in Figs. 2(a) ~ 2(d). As mentioned above, the transmission properties of a single plasmon interacting with directly coupled two QDs could be quite different with those with uncoupled QDs, which implies that the dipole-dipole interaction between the two QDs could change the transmission of a single plasmon greatly.

Figure 3 shows the transmission spectra of a single plasmon versus the interparticle distance, $kl$, with different interparticle coupling strengthes for the resonant case, $\omega_k=\Omega_1=\Omega_2$. As we can see easily from the red dashed line in Fig. 3, there appears only a complete reflection for all the interparticle distances when there is no interparticle coupling between the two QDs, $f=0$. However, as the interparticle coupling strength becomes stronger, the transmission spectrum of a single plsmon could be changed greatly. For example, when $f=0.5$, the transmission spectrum of a single plasmon is shown in Fig. 3, with the blue dash-dotted line. What is more interesting is that even though all the transition frequencies of the two QDs are resonant with the frequency of a single plasmon, there appear two complete transmission peaks, which implies that switching of a single plasmon could be achieved by adjusting the interparticle distance. When the interparticle coupling strength becomes stronger, $f=1$, there appears a single wide-band complete transmission peak. The calculations show that the interparticle coupling and distance between the two QDs are also key parameters to control the transmission of a single plasmon. Especially, we can find a graphic illustration of the coupling effect of the two

QDs in Fig. 3, in which the transmission spectrum is quite different from that of a single plasmon interacting with a single QD [18].

Next, we consider the case where the transition frequencies are different with each other. Figure 4 shows the transmission spectra of an incident single plasmon interacting with the two QDs, the transition frequencies of which are $\Omega_1 = 1.0125\, \omega_0$ and $\Omega_2 = 1.0129\, \omega_0$, respectively. When the interparticle coupling strength is weak, $f=0.1$[Fig. 4(a)], the single plasmon is completely reflected at resonances, $\omega_k=\Omega_1$ or $\omega_k=\Omega_2$, and the two-QD system behaves as a mirror, which means the transmission of the single plasmon can be switched on or off by dynamically tuning the transition energies of the two QDs. When the interparticle coupling strength is strong, $f=1$, the plasmon is not completely reflected at resonances, but at a little different frequencies, as shown in Fig. 4(b). We also found that the order of the QDs does not influence on the transmission of the incident plasmon at all. For the given transition energies $\Omega_1$ and $\Omega_2$, the minimum value of the transmission peak appears always at $\omega_k=(\Omega_1+\Omega_2)/2$ only when $kl=\pi/2$ as shown in Figs.4(a) and 4(b).

We can also consider the case where only a QD of the two QDs is resonant with the incident single plasmon frequency. Figure 5 shows the reflection coefficients versus the detuning $\Delta_2$ with different interparticle coupling strength $f$ for fixed interparticle distances, $kl = 0.5\pi$ [Fig. 5(a) and 5(b)] and $kl = 0.1\pi$ [Fig. 5(c) and 5(d)]. As we can see easily from the Figs. 5(a)~5(d), as the interparticle coupling strength becomes stronger, the position of the complete reflection peak moves further to the left(right) for the negative(positive) detuning $\Delta_1=-1(\Delta_1=1)$, which implies that one can control the position of the complete reflection peak by adjusting the interparticle coupling strength and the detuning. We also found that as for $kl = 0.5\pi$, the transmission curve in the case of $\Delta_1=-1$ is symmetrical to that of $\Delta_1=1$ with respect to the zero detuning, but as for $kl = 0.1\pi$, the transmission curves are asymmetrical each other. Anyway, the above result implies that one can switch the transmission properties of an incident single plasmon by adjusting the interparticle coupling strength and the detuning.

It is also interesting to consider the transmission spectrum of a propagating plasmon interacting with two QDs strongly depends on the interparticle distances, as well as the interparticle coupling strength. Figure 6 shows the transmission spectra of a single plasmon versus detuning $\Delta_2$ for various interparticle distances. When the transition frequency of one of the two QDs is resonant, $\Delta_1=0$, the transmission spectrum of a single plasmon is shown in Fig. 6(a), from which one can see that even at the resonance, $\omega_0=\Omega_1=\Omega_2$, the incident single plasmon cannot be completely reflected because of the coupling between the two QDs. We also note that all the transmission spectra have symmetrical curves with Lorentz line shape when $\Delta_1=0$, as plotted in Fig. 6(a). However, when $\Delta_1=0.3$, the transmission spectrum has asymmetrical Fano-type shape, as shown in Fig. 6(b). As shown in Fig. 1(a), the system under consideration consists of the two QDs and a metal nanowire, the energy level structure of which is shown in Fig. 1(b). In the two QDs in such a hybrid system, the excitations are the discrete interband excitons and in MNW, the excitations are the surface plasmons with a continuous spectrum. The coupling between the continuum excitations and the discrete excitations could result in the Fano-type transmission spectrum.

We address several remarks concerning the experimental realizations for the scheme proposed in this paper. At the device level, the results obtained in this paper can be utilized in such a way that two QDs can be attached to a metallic nanowire. In those schemes, quantum coherence could be generated by an incident laser beam, while the signal is launched through the nanowire as a propagating plasmon, as shown in [32]. Recently, the first experimental demonstration of exciton- plasmon coupling between a silver nanowire and a pair of QDs was reported [33]. Furthermore, the interparticle coupling effects have been attracted more and more [34, 35].

**4. Conclusions**

In summary, we investigated theoretically the transport properties of a single plasmon interacting with two QDs coupled to 1D plasmonic waveguide which is a metal nanowire via the real-space approach. The switching of a single plasmon could be controlled by adjusting the interparticle coupling strength, the interparticle distances and spectral detuning. We also show that controlling the interparticle coupling strength between the two QDs results in broadening of the band of complete reflection and distinguished shift of the complete reflection peak. The exciton-plasmon coupling between the continuum excitations in 1D plasmonic waveguide (MNW) and the discrete excitations in QDs exhibits the Fano-type transmission spectrum. Our calculation shows that the transport properties of a single plasmon interacting with two direct-coupled QDs could be influenced greatly by the interparticle coupling strength. The transport properties of a single plasmon interacting with such a two direct-coupled QDs system discussed here could find the applications in the design of the next-generation quantum devices and quantum information processing. Dissipative processes cannot be avoided in real systems. The quantum noise in nonwaveguide modes and the decoherences in the coupled system would destroy some of the interference effects found in this paper. However, the assumption of our numerical calculation is that the radiation emitted by QDs is entirely captured by the waveguide modes. We also didn't show a way to control the interparticle coupling strength. In the future, it is hoped to address the decoherence issues and the way to control the coupling effects between the QDs.

**Acknowledgments.** This work was supported by the National Program on Key Science Research of DPR of Korea (Grant No. 131-00). This work was also supported by the National Program on Key Science Research of China (2011CB922201) and the NSFC (11174229, 11204221, 11374236 and 11404410) and the Foundation of Talent Introduction of Central South University of Forestry and Technology (104-0260).

**Figure Captions**

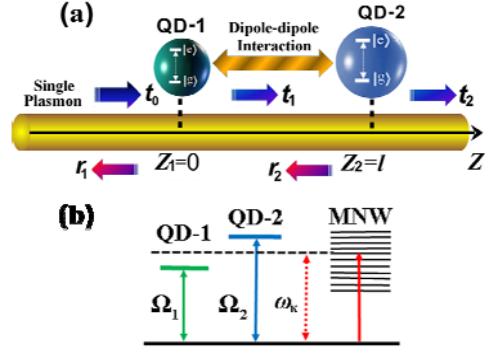

**Fig. 1** (Color online). The schematic diagram (a) and the energy level structure (b) of a nanosystem consisting of a single plasmon and two direct-coupled QDs System coupled to 1D waveguide. $t_i$ and $r_i$ are the transmission and reflection amplitudes at the place $z_i$, respectively.

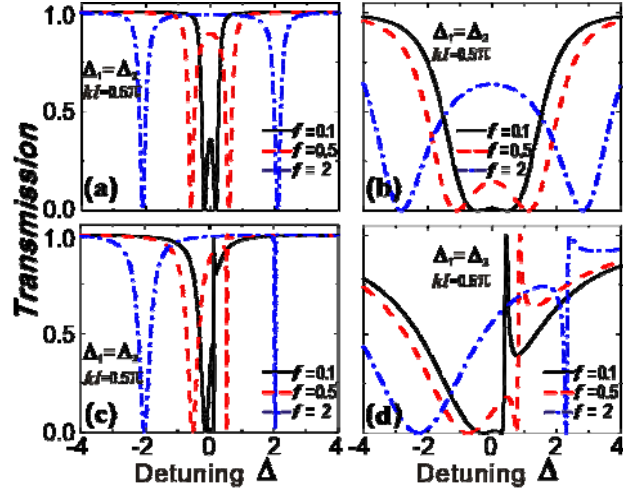

**Fig. 2** (Color online). The transmission spectra of a propagating single plasmon interacting with two direct coupled QDs versus detuning $\Delta$, where $f=0.1$ (solid line), $f=0.5$ (dashed line), and $f=2$ (dash-dotted line). (a) $J=0.1$, $kl = 0.5\pi$ (b) $J=1$, $kl = 0.5\pi$ (c) $J=0.1$, $kl = 0.1\pi$ and (d) $J=1$, $kl = 0.1\pi$. Here we set $\omega_0 \equiv 2\pi v_g / l$ and the units of $J$ and $f$ are $10^{-4}\omega_0$, while the unit of detuning $\Delta$ is $10^{-3}\omega_0$. In all cases two QDs have the same transition energies($\Delta_i = \Delta$ ($i = 1, 2$)).

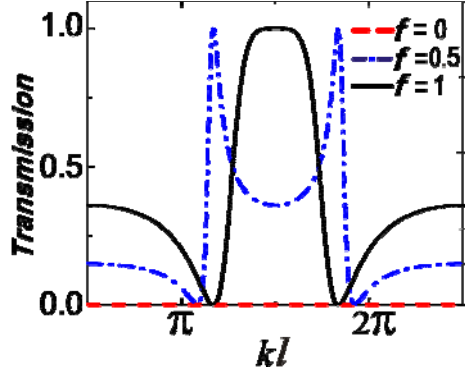

**Fig. 3**(Color online). The transmission spectra of a propagating single plasmon interacting with two direct coupled QDs versus interparticle distance, $kl$, where $f=0$ (dashed line), $f=0.5$ (dash-dotted line), and $f=1$ (solid line). Here we set $\omega_0 \equiv 2\pi v_g / l$, $J=1$ and the units of $J$ and $f$ are $10^{-4}\omega_0$. In all cases two QDs have the same transition energies, resonant with the frequency of the incident single plasmon, $\Delta_i = 0$ ($i = 1, 2$).

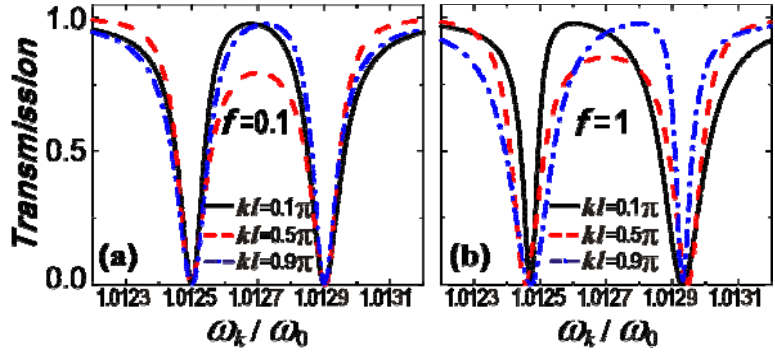

**Fig. 4**(Color online). The transmission spectra of a propagating single plasmon interacting with two direct coupled QDs versus the incident frequency $\omega_k$, where $kl = 0.1\pi$ (solid line), $kl = 0.5\pi$ (dashed line), and $kl = 0.9\pi$ (dash-dotted line). (a) $f=0.1$, (b) $f=1$. The two QDs have the different transition energies ($\Omega_1 = 1.0125\,\omega_0$, $\Omega_2 = 1.0129\,\omega_0$). Here we set $\omega_0 \equiv 2\pi v_g / l$, $J=1$ and the units of $J$ and $f$ are $10^{-4}\omega_0$.

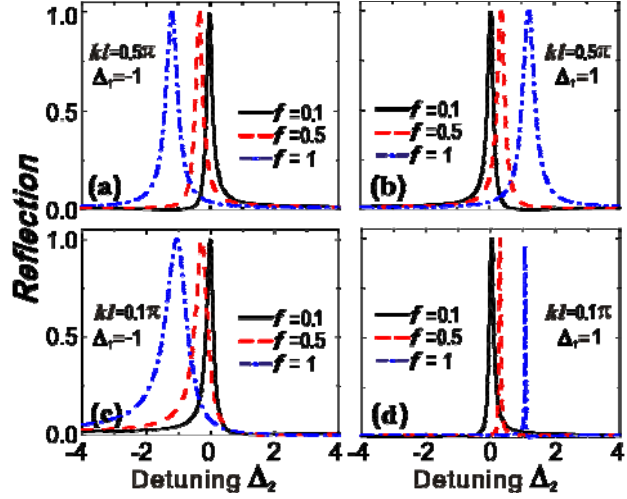

**Fig. 5**(Color online). The transmission spectra of a propagating single plasmon interacting with two direct coupled QDs versus detuning $\Delta_2$, where $f=0.1$ (solid line), $f=0.5$ (dashed line), and $f=1$ (dash-dotted line). (a) $\Delta_1=-1$, $kl = 0.5\pi$ (b) $\Delta_1=1$, $kl = 0.5\pi$ (c) $\Delta_1=-1$, $kl = 0.1\pi$ and (d) $\Delta_1=1$, $kl = 0.1\pi$. Here we set $\omega_0 \equiv 2\pi v_g / l$, $J=0.1$ and the units of $J$ and $f$ are $10^{-4}\omega_0$, while the unit of detuning $\Delta_i$ ($i = 1, 2$) is $10^{-3}\omega_0$.

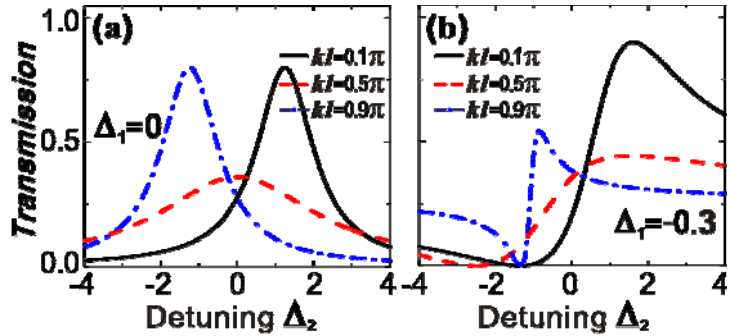

**Fig. 6**(Color online). The transmission spectra of a propagating single plasmon interacting with two direct coupled QDs versus detuning $\Delta_2$, where $kl = 0.1\pi$ (solid line), $kl = 0.5\pi$ (dashed line), and $kl = 0.9\pi$ (dash-dotted line). (a) $\Delta_1=0$, (b) $\Delta_1=-0.3$. Here we set $\omega_0 \equiv 2\pi v_g / l$, $J=0.5$, $f=0.5$ and the units of $J$ and $f$ are $10^{-4}\omega_0$, while the unit of detuning $\Delta_i$ ($i = 1, 2$) is $10^{-3}\omega_0$.